\documentclass[twocolumn,aps,prl,reprint,groupedaddress,longbibliography]{revtex4-2}

\usepackage[utf8]{inputenc}
\usepackage{graphicx}% Include figure files
\usepackage{dcolumn}% Align table columns on decimal point
\usepackage{bm}% bold math
\usepackage{float}
\usepackage{mathtools}
\usepackage{CJK}
\usepackage{color}
\usepackage{amsmath}
\usepackage{hyperref}
\usepackage{color,soul}

\begin{document}
\begin{CJK*}{GB}{}

\title{Model for the dynamics of  carrier injection in a band with polaronic states: application to the exciton dissociation in Organic Solar Cells }

\author{$~~~~~~~~~~~~~~~~~~~~~~Khouloud ~Chika^1, Alexandre ~ Perrin^{2}, Jouda ~ Jemaa ~ Khabthani^1, Ghassen~Jema\Ddot{\textrm{i}}^{1},\newline Jean\textrm{-}Pierre ~ Julien^{2}, Samia~Charfi~Kaddour^1, Didier~Mayou^{2}$}

\affiliation{$^1 Universit \acute e~Tunis~El~Manar,~Facult \acute e ~des~Sciences~de~Tunis,~ Laboratoire~de~Physique~de~la~Mati\grave e re~Condens\acute ee,~1060~Tunis,~Tunisia.~~ \\
^2 Universit\acute e~ Grenoble~ Alpes,~ CNRS,~ Institut~ NEEL,~ F-38042~ Grenoble, ~France$}

\collaboration{E-mail:  khouloud.chika@fst.utm.tn}

\begin{abstract}
      We develop a quantum model for the dynamics of carrier injection in a band that presents a strong carrier-vibration coupling. This coupling modifies the spectral density of the band and can even create pseudo-gaps that sign the onset of polaronic states. The injection of a carrier that interacts with many vibration modes is a complex many-body process that is treated by combining the quantum scattering theory and the Dynamical Mean-Field Theory (DMFT). For the model analysed here, which is adapted to compact phases, the number Z of neighbors of a given site is large and in this limit the DMFT becomes exact. The model is applied to the  excitonic dissociation at the donor-acceptor interface for organic solar cells. The main ingredients are the electron-hole Coulomb interaction, the recombination process and the existence of polaronic states in the acceptor band. Using parameters extracted from ab-initio calculations we analyze the spectral density on the charge transfer state (CTS), the average energy transfered to phonons on the CTS and the quantum yield of the injection process. We find in particular that, even with a strong electron-vibration coupling, one can get a vibrationally cold charge transfer state with a high injection yield as often observed experimentally.

\end{abstract}
\maketitle
\end{CJK*}

The interfacial charge transfer (CT) between heterogeneous materials constitutes a key physical phenomenon central to a variety of light-induced energy transport and conversion processes such as photocatalysis, photovoltaics, energy storage, molecular electronics, etc... \cite{xue2021interfacial,yan2019efficient,wang2021interfacial, fujisawa2020interfacial,uddin2022organic,nayak2019photovoltaic,zhang2020establishing,gao2014charge}. In the case  of excitonic solar cells such as heterojunction bulk organic solar cells (OSCs) the photon is absorbed in the donor zone and  leads to the creation of an exciton which is stable because of the strong Coulomb interaction between the electron and the hole constituting it. The exciton must migrate to the donor-acceptor interface in order to dissociate \cite{etxebarria2015polymer,karki2021path}. Yet once the charges are in separate phases, they still need to overcome their mutual Coulomb attraction which is larger than the thermal energy room (around 0.025 eV) otherwise they recombine \cite{deibel2010polymer,ono2018origin,street2010interface,gohler2018nongeminate,zhu2009charge}. 
Several phenomena have been identified as facilitators of charge separation such as built-in electric fields at donor-acceptor interfaces, delocalization of the excitons and of the free carriers charges,  the offset  in energy levels between donor and acceptor, structural disorder, etc...\cite{clarke2010charge,baranovskii2012calculating,nayak2013separating,bassler2015hot,few2015models,d2016charges,brey2021quantum}. A central concept is that of the charge transfer state (CTS) which is the first state in which the electron hops on the acceptor side and on which it is coupled to local vibration modes.  Because of this coupling, the electron can excite one or several phonons of the vibration mode which corresponds to a vibrationally hot CTS. If no phonons are excited this corresponds to a vibrationally cold CTS.  A much-debated issue is how the efficiency of the charge separation is related to the release of energy on these modes (vibrationally hot CTS) or not (vibrationally cold CTS) \cite{bassler2015hot,gautam2016charge,lee2022intrachain,grancini2013hot,zheng2017charge,antropov1993phonons,castet2014charge,d2016electrostatic,bera2015impact}.

On the theoretical side, the treatment of charge transfer in organic semiconductors is complex. Historically some phenomenological models were developed such as the Braun-Onsager \cite{onsager1934deviations,onsager1938initial} analytic model which is based on a classical picture that allows to describe separation in a strong  Coulomb potential. The Marcus theory and related approaches are also much-used and describe charge transfer at the molecular level with incoherent hopping \cite{marcus1956theory,bassler2015hot,few2015models,fazzi2017hot,chen2018assessing,tscheuschner2015combined,athanasopoulos2017efficient,bassler2015hot,few2015models,fazzi2017hot,chen2018assessing,tscheuschner2015combined}.
Ab-initio electronic structure calculations \cite{fujita2018thousand,d2016charges,brey2021quantum} also bring much useful information concerning the electronic states and the electrostatic potential. For a fully microscopic understanding of the charge separation mechanism in these organic photovoltaic devices, numerical methods have been proposed such as exact diagonalization \cite{wellein1997polaron} and time-dependent density functional theory \cite{ku2011time,andrea2013quantum,polkehn2018quantum}. Yet the problem of describing properly the charge separation process is still largely open and new complementary approaches are needed.

In the present study, we develop a formalism for the analysis of carrier injection in a band with polaronic states. The injection of a carrier that interacts with many vibration modes is a complex many-body process that is treated by combining the quantum scattering theory and the Dynamical Mean-Field Theory (DMFT). This avoids perturbative theories, such as the Fermi Golden Rule \cite{wilcox2015ultrafast,zhao2012charge}, which is often used but is justified only at the weak electron-vibration coupling. This mean-field approach becomes exact  for models with a number Z of neighbors which becomes large and is therefore well adapted to compact systems. We apply it to the dynamics of the exciton dissociation process in organic solar cells. For simplicity, we consider only one band on the acceptor side  and retain only the excitonic state on the donor side which couples most efficiently to charge transporting states \cite{bassler2015hot}. We use a simple Holstein Hamiltonian which describes an electron that interacts with one mode on each site of the acceptor. All the modes are equivalent and have the same frequency and the same coupling to the electron. In addition, we include  the electrostatic potential  due to the electron-hole interaction near the donor-acceptor interface. On the acceptor side,  we do not include a disorder potential whose effect on the injection process can be moderate as shown in \cite{vazquez2013calculation}. A typical range for the parameters of the Hamiltonian is estimated from ab-initio calculations \cite{antropov1993phonons,faber2011electron,castet2014charge,d2016electrostatic,zheng2017charge}. We analyze the spectral density on the CTS, the energy released by emission of phonons on the CTS and the quantum yield. We establish in  particular a phase diagram for the energy released on the CTS and the quantum yield as a function of the injection energy and of the recombination rate. We find that in the presence of the electron-hole attraction, the electron can be injected into the acceptor with  moderate initial energy and we show that the charge transfer state can stay vibrationally  cold \cite{bassler2015hot,gautam2016charge} with a high yield even with a strong electron-vibration coupling.  

\begin{figure}[h!]
\includegraphics[height=4.0cm,width=8cm]{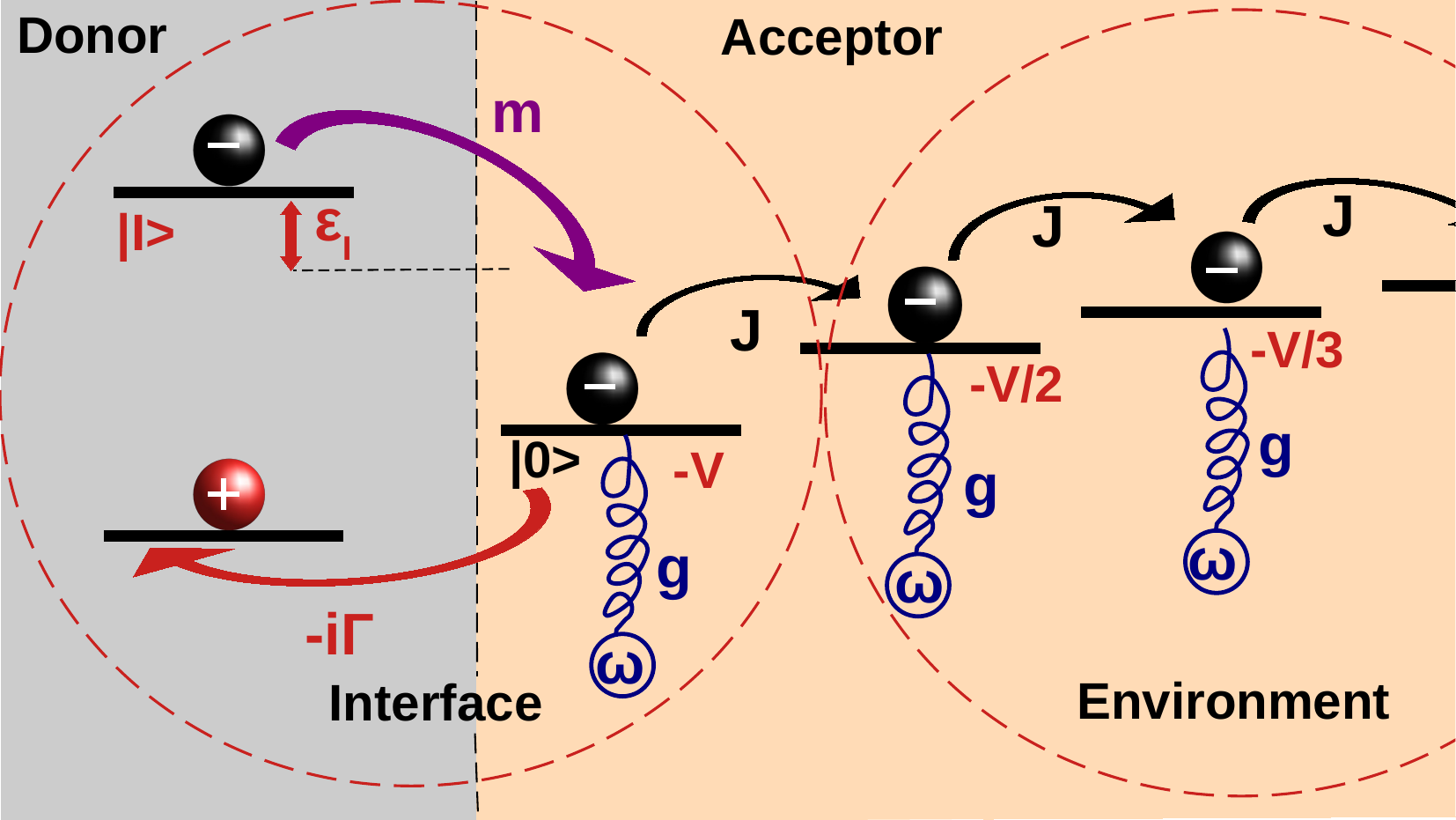}
\caption{\label{fig1} Representation of the model. $|I>$ is the initial state when the electron is on the LUMO orbital on the donor side. $|0 >$ is the state when the electron is on the CTS (L=0)  where it can recombine with a rate $\Gamma$. The electron propagates through the different layers L=0,1,2.. and can excite the local vibration modes. }
\label{Model}
\end{figure}

Our charge separation model is presented in Fig. (\ref{Model}). We consider optical modes with frequencies higher than the thermal energy at room temperature so that all vibration modes are initially empty when the charge injection process starts. The red ball is the hole that is considered fixed and the black balls represent the different positions of the electron. The state $|I>$ is the excited singlet state from which the electron can be injected into the  charge transfer state (CTS) via the hopping integral $m$. In this state $|I>$, the electron is on the LUMO orbital on the donor side and the hole is on the HOMO orbital. The electron can  jump, on the sites $i$ of the successive layers $L=0,1,2... $  (the layer $L=0$ is the CTS itself) and it can excite phonons on all sites of a given layer. We assume that the sites $i$ are distributed on a Bethe lattice \cite{haydock1980recursive,katsura1974bethe} which is known to reproduce correctly the local environment of a compact system (see Supplementary Material (SM) \cite{Supplemental}). For simplicity, we take the standard limit of infinite coordination of this Bethe lattice, in which case the mean-field solution given by the DMFT is exact.\\

In this letter the total Hamiltonian $H$ takes the form  \cite{holstein1959studies,holstein1959studiess}: \\
\begin{equation}
\begin{aligned}
H=& \varepsilon_{I} c_{I}^{+} c_{I}-m(c_{I}^{+} c_{0}+c_{0}^{+} c_{I})-\sum_{i} \frac{V}{L_{i}+1}c_{i}^{+} c_{i}\\
& +\sum_{i}\hbar\omega a_{i}^{+} a_{i}-\sum_{i,j} J_{i,j}\left(c_{i}^{+} c_{j}+c_{j}^{+} c_{i}\right)\\
&+\sum_{i} g_i c_{i}^{+} c_{i}\left(a_{i}^{+}+a_{i}\right) + H_R
\end{aligned}
\label{eq1}
\end{equation}

 Where the index $i$ denotes the different sites, $i=0$ being the CTS. $\varepsilon_{I}$ is the energy of the incoming electron of a molecule at the donor site. The hopping parameter $m$ between the donor site and the CTS will be taken as weak compared to the pure electronic bandwidth $4J$ so that we can take the limit  $m\to$ 0. For a site $i$ the electron creation (annihilation) operators are $c_{i}^{+}\left(c_{i}\right)$. The electrostatic potential at site $i$ which is on the layer $L_i$ is $-\frac{V}{L_{i}+1}$. It is  due to the electron-hole interaction and is characterized by the parameter $V$. The phonon creation (annihilation) operators of the local vibration mode at site $i$ is $a_{i}^{+}\left(a_{i}\right)$ and $\hbar \omega $ is the phonon energy. $J_{i,j}$ are the hopping matrix elements between nearest neighbors $i$ and $j$ on the Bethe lattice and are expressed from $J$ as explained in the SM. $g$ is the electron-vibration coupling parameter. The Hamiltonian $H_R$ represents the recombination processes. 

\begin{figure}[h!]
\includegraphics[height=4.0cm,width=8cm]{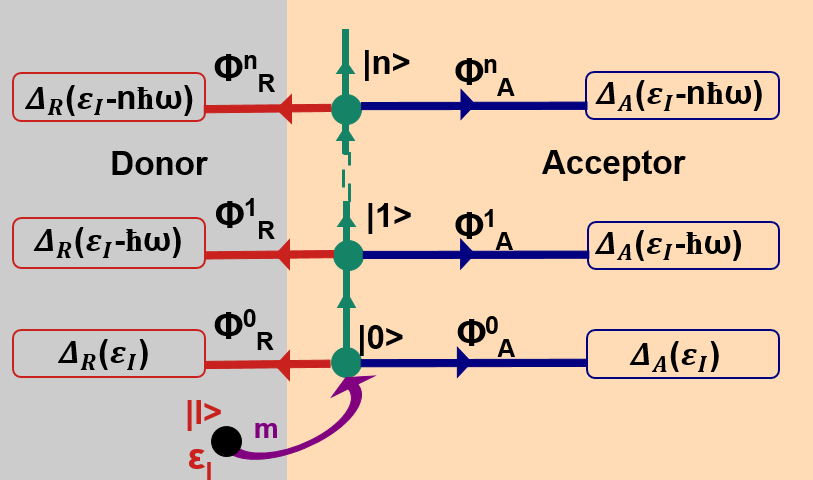}% Here is how to import EPS art
\caption{\label{fig2} Schematic representation for the charge injection process in the Hilbert space. $\Phi_R^n$ are the recombination fluxes that enter the recombination channels and $\Phi_A^n$ are the injection fluxes that enter in the injection channels (acceptor side). The values of the fluxes are determined from the self-energies $\Delta_R(\varepsilon_I-n\hbar\omega)$ and $\Delta_A(\varepsilon_I-n\hbar\omega)$ as explained in the SM}
\label{Hilbert}
\end{figure}

The injection process is analyzed in the full Hilbert space of the electron + vibration modes system (represented in Fig. (\ref{Hilbert})). The initial state $|I>$ corresponds to the electron on the donor side with no vibration mode excited. The hopping term $m$ allows the transfer to the CTS with zero phonons $|0>$. Then the electron-vibration coupling couples the states $|0>,|1>,...|n>$ with $n$ phonons created on the CTS and no other mode excited on the acceptor side. Starting from state $|n>$ there are two channels in which the electron leaves the CTS. Channel $nA$ corresponds to the propagation on the acceptor side and channel $nR$ corresponds to the recombination process. The probabilities of injection in channels $nA$ ($nR$) are $\Phi_{A}^n$ ($\Phi_{R}^n$). From these quantities, it is possible to express the quantum yield $Y$ which is the probability for the electron to be injected into the acceptor and the average vibration energy $E_{TS}$ on the CTS \cite{nemati2016modeling}. One has :

\begin{subequations}
\begin{flalign}
Y=\sum_{n=0}^{\infty} \Phi_{A}^{n}  =1-\sum_{n=0}^{\infty} \Phi_{R}^{n} \\
E_{TS}= \sum_{n} n\hbar\omega [ \Phi_{A}^{n}+ \Phi_{R}^{n}]
\end{flalign}
\label{Def}
\end{subequations}

The probabilities $\Phi_{A}^n$ and $\Phi_{R}^n$ of injection in the channels $nA$ and $nR$  are computed from the scattering theory \cite{wellein1997polaron,wellein1998self,ciuchi1997dynamical,de1997dynamical,richler2018inhomogeneous} and depend only on the self-energies $\Delta_A (\varepsilon_I -n\hbar\omega)$ and $\Delta_R(\varepsilon_I-n\hbar\omega)$ which represent the effect of the corresponding channels. As shown in SM $\Delta_A(z)$ is computed from the DMFT.

Solving the coupled mean-field equations for an inhomogeneous model, as for the case with electron-hole interaction, is difficult with standard methods. In order to solve these equations, we use  development in continued fractions  which is very efficient \cite{richler2018inhomogeneous}. In this approach, the self-energy $\Delta_A(z)$ is represented by a continued fraction with a finite number of stages N (N is up to 2000) and the energy resolution improves when N increases. Mathematically the real and imaginary parts of  $\Delta_A(z)$ are related so that one has only to reconstruct the imaginary part. For a finite continued fraction with $N$ stages  the imaginary part is a sum of $N$ delta peaks with $Im\Delta_A(z)=\sum p_{i}\delta(E-E_{i})$. In order to limit numerical fluctuations, we convolute each delta peak by a Gaussian function with an adjustable width $\Delta E$ which determines the energy resolution (see SM for more details). Physically such a broadening can be seen as reproducing a small coupling to other degrees of freedom not considered in the Hamiltonian. Here the resolution is estimated in the range 0.1-0.2 J which is sufficient for the present discussion.

The recombination process is modeled as an injection into a continuum and we discuss two limits where this continuum is either narrow with a width comparable to the electronic acceptor  band (of the order of $4J$) or is much wider. We expect that the wide band limit is valid if the hole and the electron recombine by emission of a photon which energy is larger than $4J$. This wide band limit can be a good approximation for other recombination pathways since the total energy of a few eV that is released is again larger than the typical bandwidth of the acceptor $4J$. For a recombination process into a wide continuum,  we neglect the dependence of $\Delta_R(z)$ with $z$ and for simplicity, we take $\Delta_R(z)=-i\hbar \Gamma$ where $\Gamma$ is the recombination rate.

We discuss now the results obtained for the LDOS on the site $|0>$, the energy $E_{TS}$ lost by the electron on the CTS and the quantum yield $Y$ of the injection. All energies are given in unit of $J$ (the pure electronic bandwidth is $4J$) and for all cases, we consider $\hbar\omega=1$. The LDOS are given in units of $1/J$. We consider four cases that combine $g=1$ or $g=\sqrt{2}$ for the electron-vibration coupling parameter with $V=0$ or $V=1.5$ for the electrostatic potential parameter (see the expression of the Hamiltonian in equation \ref{eq1}). Additional results for other values of the parameters confirm the behavior discussed here (see SM). 

\begin{figure}[h!]
\includegraphics[height=3.9cm,width=8cm]{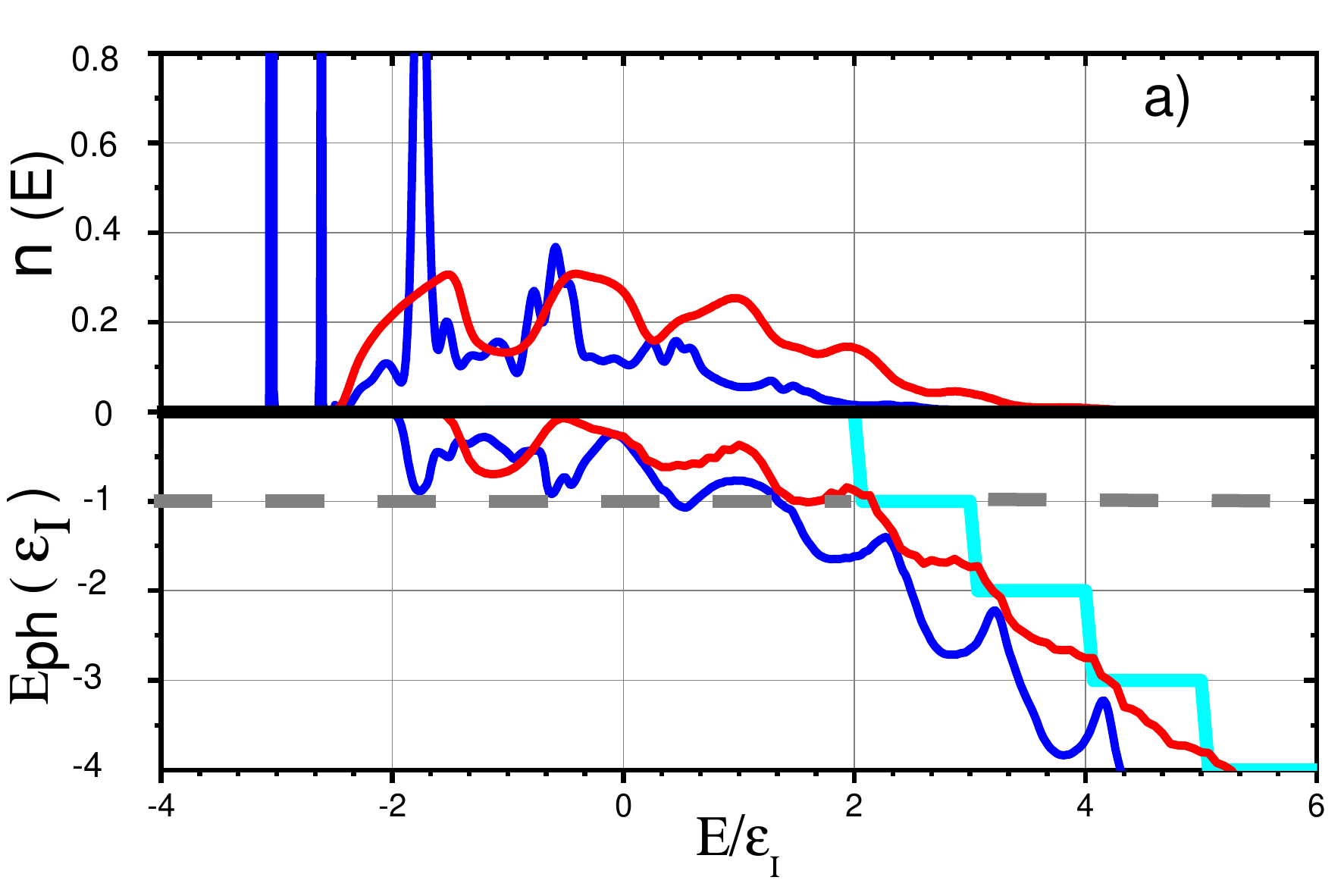}
\includegraphics[height=3.9cm,width=8cm]{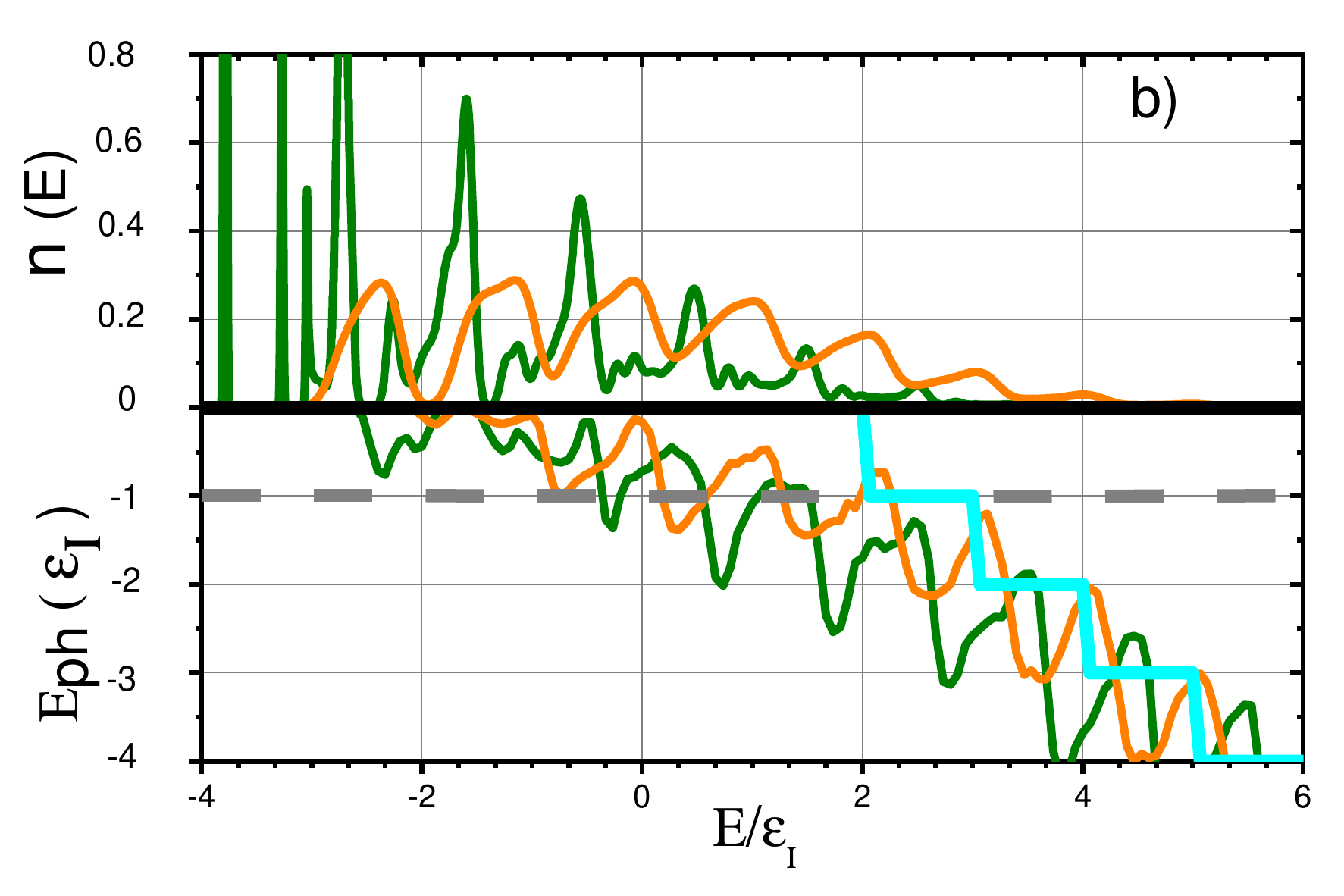}
\caption{\label{fig3}  The spectral density $n (E)$ and the average energy lost on the CTS $E_{ph}(\varepsilon_I)$ ($E_{ph}=-E_{TS}$) without recombination. The cyan curve represents $E_{ph}(\varepsilon_I)$ in the case where g$\to$ 0.  For panel (a) $g=1$ and the potential parameter is  $V=0$ (red) and $V=1.5$ (blue). For panel (b)  g=$\sqrt{2}$ and the potential parameter is  $V=0$ (orange) and $V=1.5$ (green). For both panels, the horizontal dashed gray line indicates the limit of vibrationally cold and hot CTS regimes.  }
\label{PureInj}
\end{figure}

Figure (\ref{PureInj}) presents results in the absence of recombination. We consider first the upper part of the panels $a)$ and $b)$ which show the LDOS $n(E)$ on-site $|0>$. Panel $a)$ shows the results for $g=1$ and for $V=0$ (red) or $V=1.5$ (blue). We define $E_{Min}$ as the minimum energy of the spectrum in the bulk that is for $V=0$. $n(E)$ tends to zero at large energies but its spectrum is infinite. $n(E)$ presents oscillations with minima separated by about $\hbar\omega=1$ which indicates the pre-formation of polaronic bands. For $V=1.5$ the electrostatic potential depends on the distance to the CTS and tends to zero far from the CTS. Therefore close to the CTS the electronic density is modified but far from the CTS one expects that the system is similar to the bulk that is to the case $V=0$. For $V=1.5$ there is a continuum part of $n(E)$ which starts at the same minimum value $E_{Min}$ as for $V=0$. This is expected because the states of the bulk can propagate up to the site $|0>$ and give therefore contributions to the LDOS $n(E)$ for all energies $E>E_{Min}$. Yet the LDOS $n(E)$ on state $|0>$ is strongly modified by the electrostatic potential induced by the electron-hole interaction. In addition, there are localized states below the minimum energy $E_{Min}$ of the bulk spectrum. These states are analogous to bound electronic states in atoms or to impurity states in semiconductors and are spatially localized around the CTS. One may expect that there is an infinite series of such states close to $E_{Min}$ but their weight is too small to be detected numerically. Yet the results indicate that the total weight of the localized states (essentially the two peaks shown in panel $a)$) is about $0.34$ which means that the continuum has a weight of about $0.66$. This confirms that the electron-hole potential strongly modifies the LDOS. For the strongest electron-vibration coupling  ($g_1=\sqrt{2}$, in panel $b)$), the two lower subbands of the continuum are nearly separated by gaps for $V=0$ and the global width is larger than for $g=1$. As for panel $a)$ the introduction of the electron-hole interaction strongly modifies the LDOS both in the continuum part and by the creation of localized states. The total weight of the localized states is about $0.20$ and therefore the continuum weight is about $0.80$. Note that very close to the bottom of the continuum there is a narrow peak with a small weight of about $0.01$. We cannot discriminate if this peak is below the continuum and localized or if it belongs to the continuum, but this has no impact on our discussion.

The lower part of the panels of the Fig. (\ref{PureInj}) represents the average energy $E_{ph}=-E_{TS}$ lost by emitting phonons on the CTS. In order to inject an electron into the bulk, the initial energy $\varepsilon_I$ must be larger than the minimum energy $E_{Min}$ of the bulk spectrum. Depending on $\varepsilon_I$ two regimes occur. In the first regime there is less than one phonon emitted on average, which is a vibrationally cold transfer state regime \cite{zhang2020efficient} if $E_{Min}<\varepsilon_I<E_{Max}$ with $E_{Max}\simeq 1.5$ for $g=1$ and  $E_{Max}\simeq 1$ for $g=\sqrt{2}$. So the range of values $\varepsilon_I$ for injecting electrons with a vibrationally cold CTS is of about $4$ i.e. close to the pure electronic bandwidth. The second regime occurs at higher values of $\varepsilon_I$ when the electron can excite one or several phonons which correspond to a vibrationally hot CTS. The existence of these two regimes can be understood by considering the limit of small $g$ and the energy conservation during the injection process (see Fig. \ref{PureInj}). Indeed when $\varepsilon_I$ is between $E_{Min}=-2$ and $E_{Max}=2$ (the values of the bounds of the continuum for the pure electronic spectrum) the system does not need to emit phonons to inject an electron in the acceptor side. But when $\varepsilon_I >E_{Max}$ it is necessary to emit a phonon prior to injecting an electron in the band  and when $\varepsilon_I >3J$ it is necessary to emit two phonons to inject electrons and so on. This leads to the staircase curve in the lower part of panel a). So the average energy of phonons for $\varepsilon_I$ in the range  $[E_{Max},\infty]$ will be $E_{TS}\simeq (\varepsilon_I-E_{Max}+\hbar\omega)$. We see that the case $g=1$ is close to $g\to0$. For $g=\sqrt{2}$,  oscillations that reflect the pre-formation of polaronic bands appear but the global trend is identical.

We focus now on the effect of recombination. If the recombination is treated as an injection in a narrow continuum the two channels $nA$ and $nR$ (see Fig. \ref{Hilbert}) correspond to narrow bands and this is a  situation similar to that of the Fig. (\ref{PureInj}). As discussed previously the energy conservation imposes the existence of a vibrationally hot CTS for sufficiently large $\varepsilon_I$. For every $n$ the flux shares between $\Phi_{A}^n$ and $\Phi_{R}^n$ so that the yield $Y$ is partial. 

\begin{figure}[h!]
\includegraphics[height=3.9cm,width=8cm]{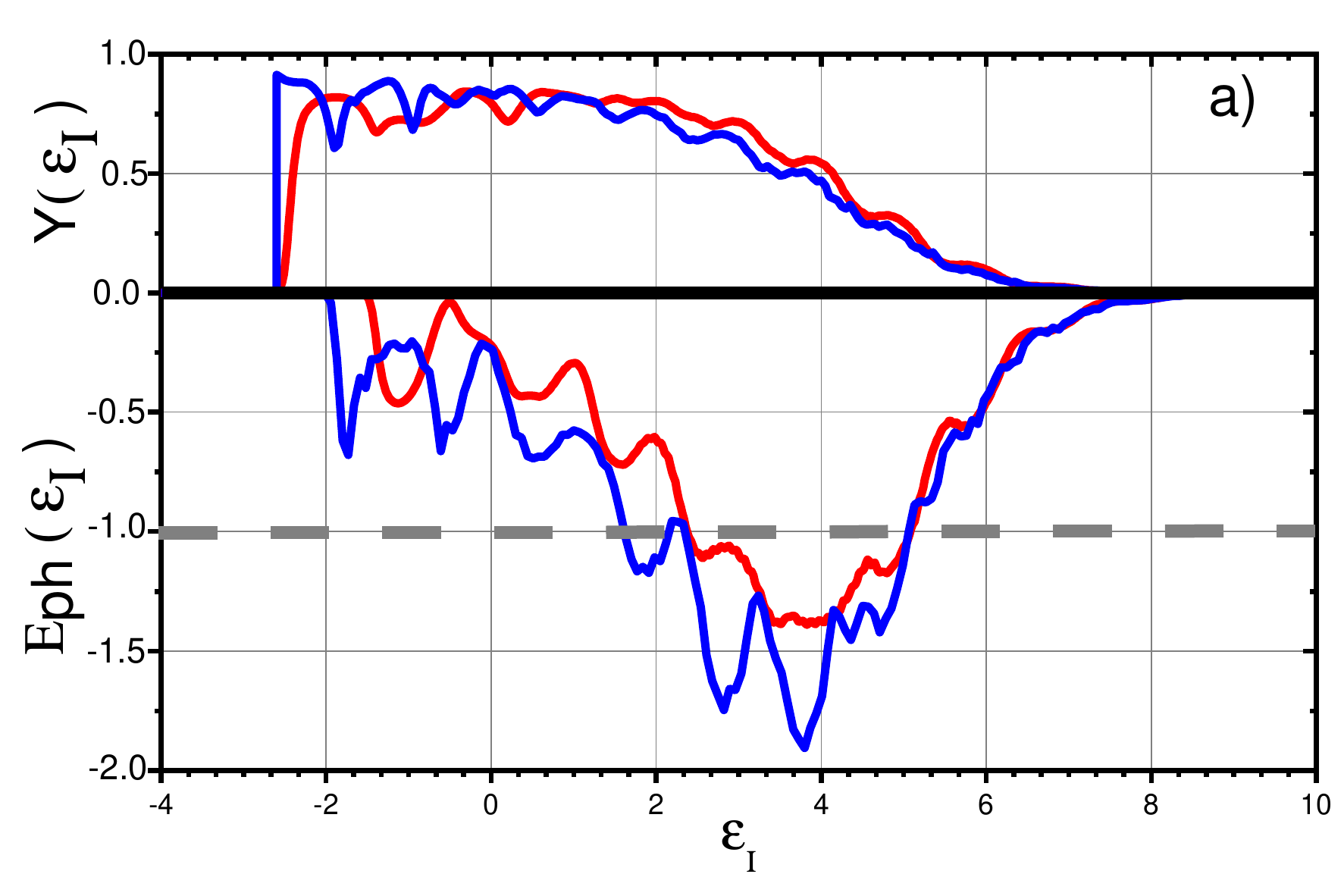}
\includegraphics[height=3.9cm,width=8cm]{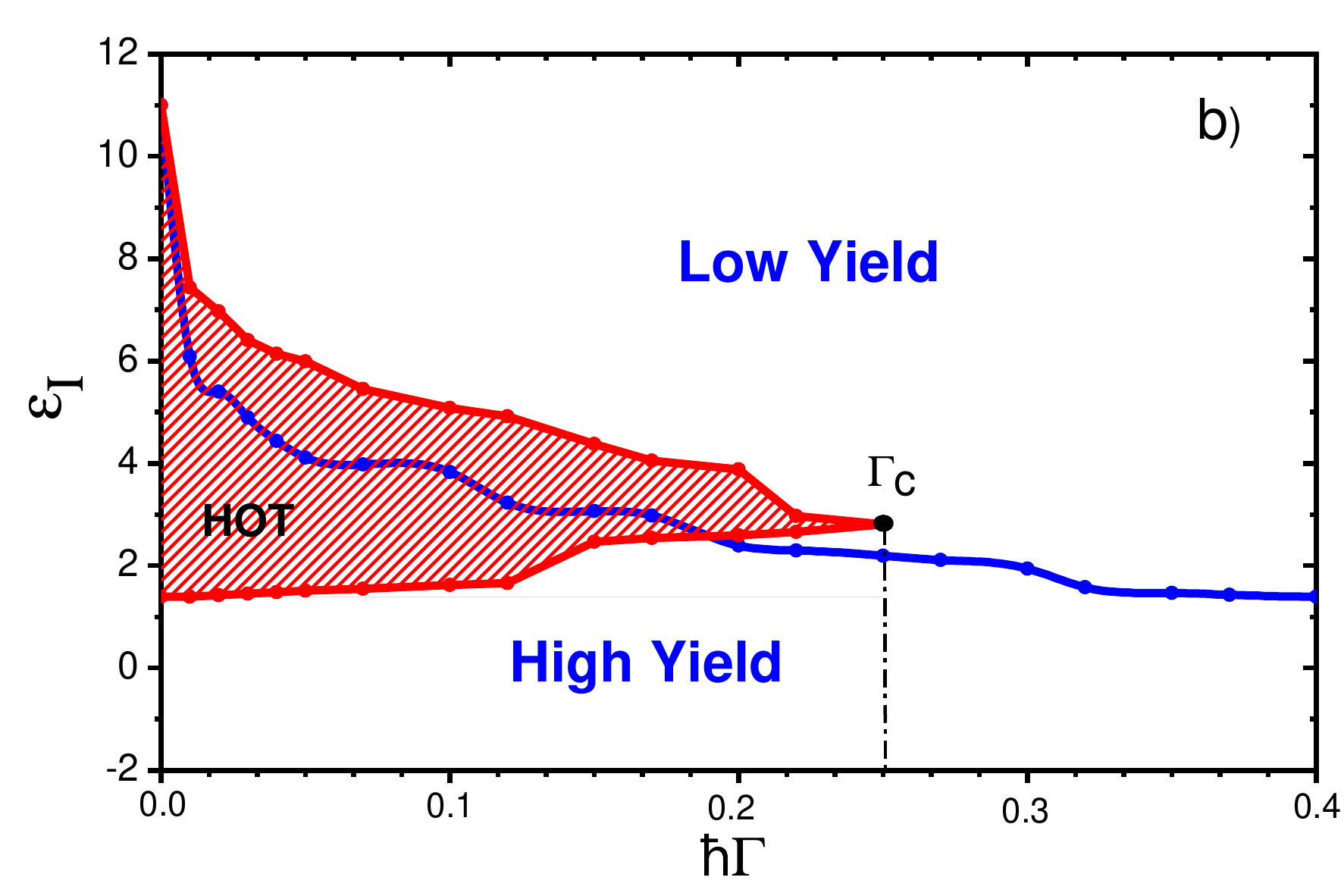}
\caption{\label{fig4} Panel $a)$ shows the quantum yield $Y(\varepsilon_I)$ and the average energy $E_{ph}(\varepsilon_I)$ lost on the CTS, with recombination $\hbar\Gamma=0.1$ and for $V=0$ (red) and $V=1.5$ (blue). The horizontal dashed gray line indicates the limit of vibrationally cold and hot CTS regimes. Panel (b) is a phase diagram as a function of the incident energy $\varepsilon_I$ and of the recombination rate $\Gamma$ for $V=1.5$. It shows the zones of high yield ($Y>0.5$) or low yield ($Y<0.5$) and the zones of vibrationally hot CTS (red) or cold CTS (white).}
\label{Interplay}
\end{figure}

The wide band limit of the recombination is completely different as we show now. We present only the case $g=1$ in Fig. (\ref{Interplay}) since results for $g=\sqrt{2}$ are qualitatively similar \cite{PRB}. The panel $a)$ shows that the results are rather similar for both values of $V=0$ and $V=1.5$. When $\varepsilon_I \leq 2$ the recombination effect is moderate and the CTS is vibrationally cold. This regime is rather similar to the vibrationally cold state regime of Fig. (\ref{PureInj}) without recombination. When $\varepsilon_I \geq 2$ the quantum yield decreases with a vibrationally hot CTS regime. For even higher values of $\varepsilon_I$ the average energy $E_{TS}$ decreases meaning again a vibrationally cold CTS regime with a small quantum yield $Y<0.5$. This results from a regime of tunneling. Indeed, the condition of energy conservation implies that the electron can enter the acceptor channels  only after having excited several phonons. A simple physical image is that the excitation of several phonons, which is performed in a tunneling regime, requires a time that increases with $\varepsilon_I$. During this time there is a transfer in the recombination channels with small $n$ at a constant rate $\Gamma$ which leads to a smaller yield and smaller number of phonons emitted. We note that if there were additional electronic bands at higher energies in the acceptor the electron could be injected into these bands without exciting phonons. This could lead to an efficient injection with a high quantum yield for high injection energy. For example, the experimental results in \cite{gautam2016charge} show that the efficiency of the injection does not depend on the injection energy. These results are not compatible with the conclusions of our model, which suggests that the injection can be done in several electronic bands in the acceptor material for these systems.
\\

In Panel $b)$ of Fig. (\ref{Interplay}) the results for $V=1.5$ are summarized in a phase diagram  with the two variables $\varepsilon_I$ and $\Gamma$. The red zone represents the vibrationally hot CTS cases with more than one phonon emitted and the white zone represents the vibrationally cold CTS regime. One sees that for sufficiently strong recombination $\Gamma > \Gamma_C$ there are no more values of $\varepsilon_I$ leading to a vibrationally hot CTS. Note that for energies between $-2$ and $E_{Min}\simeq -2.4$ (not shown here) the yield will decrease again and becomes zero if $\varepsilon_I <E_{Min}$.

Several additional comments are useful. In the presence of electron-hole interaction localized states are created for which the electron is spatially close to the CTS. Despite these localized states it is possible to inject electrons at higher energies with a high yield. Indeed for a particle that is initially in a state (the LUMO orbital of the donor side) coupled to a continuum, this corresponds to a resonant injection into a continuum \cite{vazquez2013calculation}.  Note that if the hopping integral $m$ had a finite value the localized states of the donor+acceptor system would have a finite component on the LUMO orbital by a hybridization effect. This has been discussed in \cite{Jema2022}.  In that case, the yield would be smaller than one even within the limit of vanishing recombination. Indeed the component of the initial state that is on the localized states cannot leave the interface. Therefore a too high value of the hopping integral "m" can be deleterious to the quantum yield due to the hybridization with the localized states. This may suggest a way to improve the quantum yield.

To conclude the present model shows that in a large range of incident energies $\varepsilon_I$, of the order of the pure electronic bandwidth $4J$, there can be a high quantum yield of injection and a vibrationally cold CTS despite the electron-vibration coupling. This is consistent with many experiments\cite{bassler2015hot}. Therefore a vibrationally cold transfer state with a high yield  is compatible with a strong electron vibration coupling which creates pseudo-gaps and polaronic states on the acceptor side. In addition, the model shows that the recombination process plays an  essential role in the occurrence of vibrationally hot or cold charge transfer states. For large values of $\varepsilon_I$ we show that the energy conservation imposes the injection to occur after several phonon excitations, which is a tunneling process. In this regime, the detailed characteristics of the recombination process have a strong influence on the quantum yield of the electron transfer. We emphasize that the present approach could be used to treat models with  other  characteristics, such as, for example, multiple vibration modes frequencies, multiple excitonic states or complex bands in the acceptor. It should also be useful for transfer processes at interfaces in other photovoltaic systems or even in photosynthetic systems \cite{song2021excitonic,schwarz2013role,hood2016entropy}.

\begin{CJK*}{GB}{}
\begin{acknowledgments}
We would like to thank Xavier Blase, Guy Trambly de Laissardi$\grave{\textrm{e}}$re and Sonia Haddad for stimulating discussions. The  numerical  calculations  have  been  performed at Institut N$\acute{\textrm{e}}$el, Grenoble. We thank Patrick Belmain, Computing center engineer, for computing assistance. 
A. Perrin thanks France 2030 ANR QuantForm-UGA for PhD Grant.
\end{acknowledgments}

\end{CJK*}
\bibliography{sample}

\newpage

\end{document}